\documentclass{article}
\usepackage[utf8]{inputenc}
\usepackage{xcolor}
\usepackage{hyperref}
\usepackage{amsmath, amsthm}
\usepackage{url}

\usepackage{graphicx}

\newcommand\david[1]{\textcolor{red}{#1}}

\usepackage{lipsum}

\title{TraceSecure: Towards Privacy Preserving Contact Tracing}
\author{James Bell, David Butler, Chris Hicks, Jon Crowcroft\\\small{The Alan Turing Institute\thanks{This work was supported by The Alan Turing Institute under the EPSRC grant EP/N510129/1}}\\
\small{\{jbell, dbutler, chicks\}@turing.ac.uk}}
\date{April 2020}
\begin{document}

\maketitle

\begin{abstract}
    Contact tracing is being widely employed to combat the spread of COVID-19. Many apps have been developed that allow for tracing to be done automatically based off location and interaction data generated by users. There are concerns, however, regarding the privacy and security of users data when using these apps. These concerns are paramount for users who contract the virus, as they are generally required to release all their data. Motivated by the need to protect users privacy we propose two solutions to this problem. Our first solution builds on current ``message based'' methods and our second leverages ideas from secret sharing and additively homomorphic encryption. 
\end{abstract}

\section{Introduction}

In this work we present two security and privacy conscious methods that allow for users to learn when they have been in contact with someone who is later confirmed to have COVID-19. The process of informing those who have previously come into contact with a diagnosed person is called contact tracing and has traditionally been carried out manually. Recently, many systems have been proposed to automate this process to enable a fast response to the current pandemic. These systems almost exclusively utilise smartphones by tracing individuals location data or their interactions via Bluetooth. This process is inherently intrusive for the individual. While the emphasis during the pandemic should be on stopping the spread, if users are reluctant to use contact tracing apps due to privacy concerns they will be rendered useless. In particular, adoption of a contact tracing apps must have a higher ``transmission rate'' than the virus itself in order to be effective \cite{DBLP:journals/corr/abs-2003-11511}. 

Solutions have been presented that allow for contact tracing to be done in a more secure manner. At present the main case study is the TraceTogether app  developed and deployed in Singapore. TraceTogether provides security between users, but relies on the assumption that the government is honest and trusted. The government learns everyone who is infected and is able to directly contact (by phone) those who may be at risk due to exposure. Even though the assumption of a trusted and non corrupt government is very strong we note they do not learn anything about interactions of non infected users. Consequently we feel TraceTogether provides a strong baseline of privacy. Other current solutions rely on similar assumptions, namely that there is a trusted party that can collect the tracing information and notify at risk individuals. We believe it is unlikely that solutions of this nature will be widely adopted in other countries (e.g. the US and UK) due to differing cultural views regarding the power the government should be afforded.

Some work has proposed solutions that do not require users to put complete trust in a third party \cite{DBLP:journals/corr/abs-2003-14412, DBLP:journals/corr/abs-2003-11511}, but to the best of our knowledge these all provide security at the expense of efficiency as they employ expensive cryptographic techniques. Our solutions do not rely on any such expensive techniques. Our first solution, presented in Section \ref{sec:message_protocol}, is based on sending messages between stakeholders in the system. We have two versions of this method requiring two or three separate non-colluding parties who administer the system. One of these parties is the health care provider (for example the NHS in the UK) and the others we envision as separate government run servers. We split the roles of the parties in such a way as to preserve the privacy of peoples interactions. 

Our second proposed solution, presented in Section \ref{sec:she protocol}, only requires the user to use a low-cost additively homomorphic encryption scheme to allow parties to securely send deposits of all their interactions to a server (the government) and learn if one of their contacts has subsequently been diagnosed. Here we utilise the user-user interaction to allow ``secrets'' or one time pads to be exchanged between the user devices, this allows strong privacy results without the need for the large overhead often associated with complex cryptography. As a result of our desire to treat all users, infected and non infected, equally, an interesting property of this protocol emerged. It allows the government to construct a social graph of interactions between pseudonyms. We believe this is potentially useful in the governments analysis of control measures they have implemented.

We believe our solutions provide an improved benchmark for allowing privacy preserving contact tracing to be realised by building on the already high bar set by TraceTogether. Moreover, we do not claim this work to be a complete solution to the problem, however due to its lightweight nature and strong privacy properties we believe the methods we propose could be of use in combating the spread of the virus worldwide.

\subsection{Contact Tracing via Bluetooth} \label{sec:contact tracing via bluetooth}

Our solution to providing secure contact tracing is based on users being able to exchange Bluetooth signals. With Bluetooth, the proximity of people can be estimated by the strength of the signals between their devices. In this way Bluetooth is able to more accurately determine contacts between people than GPS location data. Groups of people in buildings will have similar or the same GPS location data however their relative Bluetooth signal strengths will be reduced by internal walls, for example. 

Determining the exact implementation of Bluetooth technology is out of scope for this work. Instead we assume that there exists a mechanism for nearby phones to exchange short tokens if the devices come within 2 meters of each other, the estimated radius within which viral transmission is a significant risk.

Bluetooth Low Energy (BLE) features a ``broadcast mode", often used for advertising, which can be used to send unsolicited messages to all BLE-enabled devices within range. Within the typical operational range of 10 meters, BLE can be used to transmit a 21-byte payload and allows proximity estimation based on the Received Signal Strength (RSS). Within a metre the positioning uncertainty can be as little as a few centimetres \cite{Faragher_BLE_2014}. 

\subsection{Related Work} \label{sec:related work}

In this section we review related work. First we consider TraceTogether, the contact tracing app that is deployed in Singapore. We then look at other proposed solutions to providing privacy preserving contact tracing.

\subsubsection{Baseline: TraceTogether} \label{sec:tracetogether}

The TraceTogether app \cite{tracetogether} was developed by the Singapore Ministry of Health. It uses the idea of Bluetooth contact tracing outlined in Section \ref{sec:contact tracing via bluetooth}. When devices are ``in contact'' the devices exchange random strings as tokens, these tokens act as a UID for the device for a set period of time. When a user is diagnosed with COVID-19 the government requires them to release their list of collected tokens --- it is against the law not to cooperate \cite{tracetogether_law}. The government also has a database that links the tokens to phone numbers and identities, allowing them to directly contact those who have been in contact with the infected user. 

This method requires the state to have a high level of control over the process. In particular, there is no privacy for users who become infected. The government learns all contacts an infected user has had which is not desirable, and unlikely to be tolerated in other countries.

The app does however offer privacy for users with respect to passive snoopers and other users. This is realised by allowing the user to generate a fresh token for each new time interval. To the best of our knowledge this is every 30 minutes.

We take the TraceTogether app as a baseline and look to improve on the level of user privacy in this work. Namely our methods provide higher levels of privacy, independent of government policy, for infected users.

\subsubsection{Decentralised Privacy-Preserving Proximity Tracing (DP3T)}

In Decentralised Privacy-Preserving Proximity Tracing (DP3T) \cite{DP3T} the authors propose a decentralised design that allows for private contact tracing. This means there is no central collection of potentially sensitive user information. In contrast our set-based protocol provides the government with a pseudonymised version of the interaction graph and our message-based protocol leaks a subset of that graph containing the interactions where one party is infected. Therefore our implicit assumption is that the government cannot or does not try to de-pseudonymise this graph. Consequently, their state adversary is stronger than ours. 

Their design however has the downside that a list of infected user's broadcasts is made public. This allows users to tell which of their contacts have been infected. Our system was designed on the basis that this leakage was unacceptable.

We believe DP3T is a strong proposal for privacy preserving contact tracing. Moreover, due to our different assumptions into what attacks are acceptable and feasible we feel our work complements theirs at this early stage of research into this area. 

\subsubsection{Alternative Solutions}

Research on providing methods for privacy preserving contact tracing is fast evolving. Here we highlight three main works from the current literature.

Cho et al.~\cite{DBLP:journals/corr/abs-2003-11511} discuss and informally define privacy notions that capture the requirements of privacy preserving contact tracing. These are the privacy notions we consider our work with respect to. The  authors present methods that extend the ideas of TraceTogether to provide slightly better privacy as well as a private messaging system which provides higher levels of privacy. We note however, the authors only present a high level view of how such a system may work. We leave a discussion of their message based protocol to Section \ref{sec:message_protocol} before we present our own message based protocol. 

Berke et al.~\cite{DBLP:journals/corr/abs-2003-14412}, use  GPS location data instead of the Bluetooth method we introduced in Section \ref{sec:contact tracing via bluetooth}. Their proposal uses Private Set Intersection (PSI), a heavyweight cryptographic  technique that can securely compute the intersection of two sets. As far as we can see the authors do not report on the estimated performance or overhead of their system. They propose that intermediary steps could be taken, by removing the PSI, to speed up performance, however this will reduce the privacy of the protocol.

\section{Background}

In this section we introduce the relevant background needed for the rest of the paper. First we introduce the privacy notions outlined in \cite{DBLP:journals/corr/abs-2003-11511}. Second, we introduce additively homomorphic encryption schemes, we employ these in our set-based protocol in Section \ref{sec:she protocol}. Finally, we introduce the parameters of the problem we are considering along with the assumptions we make on Bluetooth technology.

\subsection{Privacy for Contact Tracing}\label{sec:privacy_terms}
Cho et al. \cite{DBLP:journals/corr/abs-2003-11511} introduce three notions of privacy that they deem relevant to the analysis of contact tracing protocols. Privacy from snoopers, from contacts and from the authorities are used as a framework for evaluating several existing and new approaches in their study. 

\subsubsection*{Privacy from Snoopers}
Snoopers are passive adversaries that capture communications between users and aim to learn a detailed view of their activities. The issue of privacy from snoopers has received significant attention in the broader context of location aware services \cite{v2x_pseudo_survey} and is primarily addressed with techniques that obfuscate the relationship between each user and the pseudonyms that they use when interacting with a service \cite{stajano_mix_zones}. The Singaporean TraceTogether contact tracing app provides privacy from snoopers in the form of time-limited pseudonyms which obfuscate each users interactions over any period in which more than a single pseudonym is used. 

\subsubsection*{Privacy from Contacts}
Raskar et al. \cite{raskar2020apps} report that there is a genuine fear of persecution in some communities for being identified as a carrier of the virus. Privacy from contacts concerns an adversary that is also a user of the system. Rather than just passively listening to user communications, contacts also learn whether or not they have been exposed to any risk of infection. Ideally, contacts would learn only whether they have a risk of being infected and would not learn which of their contacts had exposed them. This means that while users who only ever contact one other individual before learning their infection risk unavoidably learn the infection status of that contact, an average user should not be able to determine the source of their potential exposure. The TraceTogether app provides optimal privacy from contacts under the assumption of a trusted government. Users are directly contacted by the authorities to inform them of their infection status and so learn nothing more than the single bit which is necessary. 

\subsubsection*{Privacy from the Government}
Privacy from the government can be interpreted in a number of ways. In the most naive implementation, a government might be trusted to learn about the movements and interactions of every individual. Upon learning about an infection, the government could perform a detailed analysis of everywhere the user had been and determine every possible transmission of the virus that may have occurred. Despite placing a high degree of trust in the government, the Singaporean TraceTogether solution does offer better privacy from the government than the naive approach. The two main techniques for privacy from the government used in TraceTogether are the decentralisation of data storage and the principle of data minimisation. Firstly, user contact is recorded by each mobile device running the app and is not routinely transmitted to the government. Secondly TraceTogether does not collect any user location data. Despite these steps, the government does learn the personal identity of any user that may have risked infection. Ideally, privacy from the government would extend both to hiding all user interactions and to providing user-control over infection status reporting. 


\subsection{Additively Homomorphic Encyrption} \label{sec:additive encryption}

Fully Homomorphic Encryption (FHE) allows for arbitrary computation over encrypted data. Such schemes currently, however, have tremendous overhead and are thus not yet practical for most real world applications. While FHE allows an arbitrary number of operations, schemes permitting only limited computation over ciphertexts can be vastly more efficient. In our protocol in Section \ref{sec:she protocol}, we must test for equality of plaintexts over encrypted values. This can be accomplished efficiently using an additviely homomorpic encryption scheme. Here we show a method for computing equality over encrypted values using an additively homomorphic scheme.

Let $\mathit{enc}$ be an encryption scheme, it is said to be additive if $$\mathit{enc}(m_0) + \mathit{enc}(m_1)$$ is a valid encryption of $m_0+m_1$.

Our use case for such schemes in Section \ref{sec:she protocol} is as follows: we want to learn if two plaintext messages $m_0$ and $m_1$ are equal based on their ciphertexts, $c_0, c_1$. Using the additive property of the encryption scheme we are able to compute $\mathit{enc}(r \cdot (m_0 - m_1))$ for $r \neq 0$. If $m_0 = m_1$ then $\mathit{enc}(r \cdot (m_0 - m_1)) = 0$. 

Additive homomorphic encryption schemes have been well studied. In this work we base our analysis of overhead in Section \ref{sec:she protocol} on the Paillier encryption scheme \cite{DBLP:conf/eurocrypt/Paillier99}. 

\subsection{System Model and Assumptions}\label{sec:Assumptions}

In this section we introduce parameters we use throughout the rest of the paper. Some should be determined by medical experts, others should be determined at the time of implementation.

\begin{itemize}
    \item Let $x$ be the distance at which the virus can be transmitted from person to person.
    \item Let $s$ be the time interval people need to be at a distance $x$ or less from each other for there to be a risk that the virus is transmitted.
    \item Let $N$ be the number of days during which a person can be infectious before diagnosis, currently this is deemed to be 14 days. 
    \item Let $t$ denote the time a users UID (a random string) is associated with their device. In TraceTogether this time period is 30 minutes.
\end{itemize}

We explicitly list the assumptions we make regarding Bluetooth technology.

\begin{itemize}
    \item Devices that are within $x$ meters of each other for time $s$ are able to transmit a 20-byte payload\footnote{We make this assumption based on \cite{herrera2016indoor}.}. 
    \item User devices are able to sample a random $uid  \in \{0,1\}^{128}$ and a random $\mathit{share} \in \{0,1\}^4$. We deem the space of UIDs to be sufficiently large for collisions not to occur, even if users generate new UIDs every 20 minutes\footnote{A city of 10,000,000 users each changing their UID every 20 minutes would only need $2^{33}$ UIDs over a 14 day period.
    }.
\end{itemize}

\section{Message Based Protocol}\label{sec:message_protocol}

The protocols we develop in this section are an extension of the ideas behind the TraceTogether app. We start by introducing and discussing the basic idea for a message based protocol as presented by Cho et al.~\cite{DBLP:journals/corr/abs-2003-11511}.

Cho et al. reduce the level of trust a user must place in the government by assuming that Alice and Bob can interact with the Government through an anonymous transport layer. The authors suggest this could be implemented using two non-colluding third parties operating a message relaying service. To realise this, however, the authors sacrifice three important properties of the above algorithm.
\begin{enumerate}
    \item Due to the anonymity of Bob, anyone wanting to spread panic could take the role of Bob and declare themselves infected when they are not, without being caught doing so.
    \item As a consequence of Alice's anonymity she must poll the server every time she wants an update on whether any of her contacts have been diagnosed. The absence of push notifications could lead to delays in notifications arriving.
    \item If the mixing service is set up especially for this purpose then whoever Alice contacts directly can learn whether she has received any updates from the size of the response.
\end{enumerate}
 
In the protocols we present here we avoid these issues by making more extensive use of the fact that we are assuming the presence of non-colluding parties.
 
We consider two users Alice and Bob. Bob is a user who may or may not be diagnosed and Alice is a user who wishes to know if any of her contacts have been diagnosed. In practice every user will play both of these roles. We call the central authority Grace (alliterative with government). In our proposed extensions we introduce Henry (a health care provider) and Mary (a messenger service) also. We split time into periods, we suggest periods of roughly 30 minutes to allow each user to change the unique identifier they transmit regularly. This is standard across all solutions of this nature, making it much harder for snoopers to track users across time by listening for their beacons.
 
 

 
 \subsection{Our Message Based Protocols} \label{sec:message based protocols}
 
 In both of the message based protocols we propose we use public key cryptography. Any semantically secure public key cryptography scheme suffices, we do not, in this section, require any extra properties such as homomorphism. Users have a fresh randomly (or pseudorandomly) generated UID for each time period for each person, we suggest taking these to be uniformly distributed in $\{0,1\}^{128}$, though any set large enough to avoid collisions will suffice.

Let the public key encryption scheme we use be denoted by $(\mathit{key\text{-}gen}, \mathit{enc}, \mathit{dec})$, $\mathit{CTR}(t,n)$ be the counter mode of a pseudorandom number generator with key $t$ and counter $n$. Let $\oplus$ denote a bitwise exclusive or operation.

\subsubsection*{Protocol 1} 

\begin{figure}
    \centering
    \includegraphics[scale=0.7]{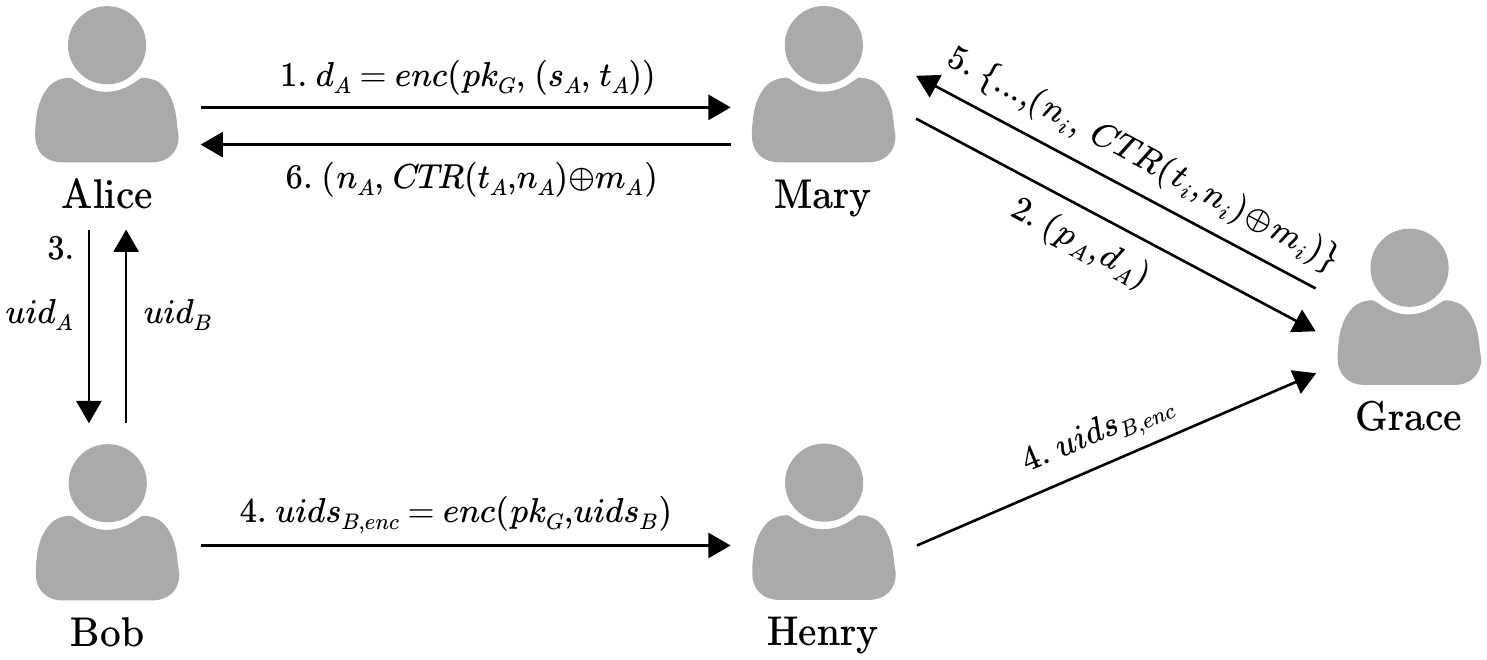}
    \caption{Overview of our message based contact tracing protocol.}
    \label{fig:message_based_protocol_1}
\end{figure}

 We introduce two extra parties. Firstly, Henry, the healthcare provider, directly confirms that Bob has COVID-19. In any normal situation this party will learn the identity of Bob and that he has COVID-19, thus we do not consider this a privacy breach. Given Henry knows Bob's identity, we will guarantee that he doesn't learn who Bob has been interacting with. This is achieved by Henry sending anonymised information to Grace. This addresses the first concern we highlight regarding the method from \cite{DBLP:journals/corr/abs-2003-11511} above. Secondly, we introduce Mary who will play the role of relaying messages between Alice and Grace thus separating Grace from the identity of Alice. This addresses the second concern as Grace can now push messages via Mary to Alice as soon as they are ready. The third concern is still a problem as Mary can see when Alice is sent a message by Grace. To address this we propose that Grace send cover traffic, at random times, to Alice even when she has not had a close contact. To Mary this will be indistinguishable from the real messages and Mary will thus not be able to tell whether Alice has received a real message. We now describe the actions of the protocol in more detail.
 

\paragraph{Setup:}
When Alice/Bob installs the app they each receive a public key belonging to Grace, $pk_G$. When Alice installs the app she randomly chooses two seeds $s_A$ and $t_A$. The seed $s_A$ will be used to generate all her future UIDs, whilst $t_A$ will be used to mask communications between her and Grace. She then sends $(s_A,t_A)$ encrypted with $pk_G$ to Mary. Mary generates a pseudonym $p_A$ records that it corresponds to Alice and forwards the pseudonym and encrypted message to Grace.

\begin{enumerate}
    \item Bob and Alice receive $pk_G$ after installing the app.
    \item Alice generates seeds $s_A$ and $t_A$.
    \item Alice sends $d_A=\mathit{enc}(pk_G,(s_A,t_A))$ to Mary.
    \item Mary assigns a pseudonym $p_A$ to Alice.
    \item Mary send $(p_A,d_A)$ to Grace.
    \item Grace decrypts $d_A$, sets $n_A=0$ and stores $(p_A,s_A,t_A,n_A)$.
\end{enumerate}

\paragraph{Interaction:}
During each time period Alice derives her UID for that period $uid_A$ form the seed $s_A$. When Bob and Alice interact they exchange UIDs. Bob maintains a set $uids_B$ of all UIDs he has seen in the last 14 days.

\begin{enumerate}
    \item At the beginning of time period $j$ Alice generates $uid_A=\mathit{CTR}(s_A,j)$.
    \item Alice broadcasts $uid_i$ throughout the time period using BLE.
    \item If Bob comes into proximity of this broadcast he stores $uid_A$ together with a timestamp for 14 days in a set $uids_B$
\end{enumerate}

\paragraph{Inform:}
When Bob is diagnosed, he gives Henry an encryption under $pk_G$ of the current $uids_B$. Henry then passes this on to Grace who decrypts them and finds the pseudonym $p_i$ corresponding to each one. For each $p_i$ present (with duplicates removed) Grace prepares a message, $m_i$, informing the user of a close contact (possibly with some indicator of the extent of the contact). She then masks these messages with a mask generated from $t_i$ and sends them labelled with the $p_i$ to Mary. From the pseudonyms Mary can tell where each message should be forwarded to and Alice receives her message.

\begin{enumerate}
    \item Bob computes $uids_{B, enc} = enc(pk_G, uids_B)$ and sends this to Henry, who passes it to Grace.
    \item Grace decrypts the message to find $uids_B$ and looks up the corresponding $p_i$.
    \item For each distinct $p_i$, Grace prepares a message $m_i$, increments $n_i$, sets $M_i = (n_i, \mathit{CTR}(t_i,n_i) \oplus m_i)$ and sends $(p_i,M_i)$ to Mary.
    \item Mary forwards $M_i$ to the user $i$.
    \item Alice recovers $m_i$ and acts accordingly.
\end{enumerate}

\paragraph{Cover:}
Grace sends cover traffic to every pseudonym she currently holds on an ongoing basis. For each pseudonym, Grace will send dummy messages to Mary with each consecutive pair of messages\footnote{These messages should be separated by an independent exponentially distributed time period with some fixed mean. This ensures that the probability of seeing a cover message in the next second is independent of previous messages. Thus the cover provided is uniform.}. These messages will consist of the pseudonym and a masked dummy message. These messages will have exactly the same format as the messages Mary receives in the Inform phase.

\begin{enumerate}
    \item Grace intermittently increments $n_A$ and sends a message $$M_A=(p_A,n_A,\mathit{CTR}(t_A,n_A)\oplus \bot)$$ to Mary.
    \item Mary forwards $M_A$ to Alice.
    \item Alice recovers a $\bot$ and ignores it.
\end{enumerate}

The biggest complicating factor in the design of the above protocol is the desire to have push notifications to Alice whilst not allowing whichever party contacts Alice to know she is being informed of a close contact. If we allow Grace to learn who is receiving these notifications (but not who they had contact with to receive it) then we can manage with one fewer non-colluding party as the roles of Mary and Grace can be merged.

\subsubsection{Protocol 2}

If we relax the need for instant notification and allow say a 30 minute delay, then the alternative protocol, we present now, allows Mary's role to be merged into Henry's. The setup and interaction steps are exactly the same as for Protocol 1 except the role of Mary is now played by Henry. The inform and cover steps are now replaced by a record step and a poll step.

\paragraph{Setup:}
When Alice/Bob installs the app they each receive a public key belonging to Grace, $pk_G$. When Alice installs the app she randomly chooses two seeds $s_A$ and $t_A$. The seed $s_A$ will be used to generate all her future UIDs, whilst $t_A$ will be used to mask communications between her and Grace. She then sends $(s,t)$ encrypted with $pk_G$ to Henry. Henry generates a pseudonym $p_A$ records that it corresponds to Alice and forwards the pseudonym and encrypted message to Grace.

\begin{enumerate}
    \item Bob and Alice receive $pk_G$ after installing the app.
    \item Alice generates seeds $s_A$ and $t_A$.
    \item Alice sends $d_A=\mathit{enc(pk_G,(s_A,t_A))}$ to Henry.
    \item Henry assigns a pseudonym $p_A$ to Alice.
    \item Henry send $(p_A,d_A)$ to Grace.
    \item Grace decrypts $d_A$, sets $n_A=0$ and stores $(p_A,s_A,t_A,n_A)$
\end{enumerate}

\paragraph{Interaction:}
During each time period Alice derives her UID for that period $uid_A$ form the seed $s_A$. She broadcasts it around her and when she comes into Bob's proximity Bob records it. Bob maintains a set $uids_B$ of all UIDs he has seen in the last 14 days.

\begin{enumerate}
    \item At the beginning of time period $j$ Alice generates $uid_A=\mathit{CTR}(s_A,j)$.
    \item Alice broadcasts $uid_i$ throughout the time period using BLE.
    \item If Bob comes into proximity of this broadcast he stores $uid_A$ together with a timestamp for 14 days in a set $uids_B$
\end{enumerate}

\paragraph{Record:}
When Bob is diagnosed, he gives Henry an encryption under $pk_G$ of the current $uids_B$. Henry then passes this on to Grace who decrypts them and finds the pseudonym $p_i$ corresponding to each one. For each $p_i$ present (with duplicates removed) Grace then prepares a message, $m_i$, informing them they have been in a close contact (possibly with some indicator of the extent of the contact). She then masks these messages with a mask generated from $t_i$ and stores them labelled with the $p_i$ ready to be sent to Henry.

\begin{enumerate}
    \item Bob computes $uids_{B, enc} = enc(pk_G, uids_B)$ and sends this to Henry, who passes it to Grace.
    \item Grace decrypts the message to find $uids_B$ and looks up the corresponding $p_i$.
    \item For each distinct $p_i$, Grace prepares a message $m_i$, increments $n_i$, sets $M_i = (p_i,n_i, \mathit{CTR}(t_i,n_i) \oplus m_i)$ and stores $M_i$ to be sent later.
\end{enumerate}

\paragraph{Sending:}
Periodically (say every 30 minutes), Grace send a message to Alice. If there is a message stored in a recent record phase but not yet sent then Grace sends this, otherwise Grace sends an encryption of a dummy message.

\begin{itemize}
    \item Every 30 minutes, if there is a stored message $M_A$ Grace sends this to Henry. Otherwise she increments $n_A$ and sends a message $$(p_A,n_A, \mathit{CTR}(t_A,n_A) \oplus \bot)$$.
\end{itemize}


\subsection{Privacy of our Message Based Protocols} \label{sec:privacy of messaged based}

\subsubsection{Privacy from Snoopers}

In both protocols users change UID to something uncorrelated between every time period it would thus be hard for a snooper to track individuals by the UIDs they broadcast. Therefore we have privacy from snoopers in both protocols.

\subsubsection{Privacy from Contacts}

In the interaction phase contacts learn nothing more than snoopers therefore this phase alone cannot compromise privacy from contacts. The only other information that users receive is in the form of messages that are sent to Alice in the inform and sending phases. Alice is not informed by the system who triggered the sending of this message. We could imagine that Alice is aware that she has only interacted with Bob in the last fortnight or knows that Bob has just received test results, in this case information is leaked to Alice about Bob. However this is also leaked by any protocol that informs Alice in a timely manner so we consider this to be an acceptable amount of information to leak.

\subsubsection{Privacy from Government}

For privacy from the government we first consider the case of no collusion. Starting with Protocol 1 in which:
\begin{itemize}
    \item Henry learns who he has diagnosed and how many UIDs they have collected. The first is not a breach as he knows this anyway, the second might be considered a problem but could be fixed by getting Bob to generate lots of dummy UIDs to submit. This is a cost that would only be paid upon being diagnosed and would prevent Henry from learning the number of UIDs submitted.
    \item Grace gets a psuedonym for each user and is able to tell which of them have been in contact with the same diagnosed people, how much and when. We argue that this could be acceptable as breaking the pseudonymity can be argued to be sufficiently hard so as not to happen.
    \item Mary learns who is using the app. On top of that Mary learns when people are receiving messages. This leakage will be negligible provided the cover traffic is intense enough compared to the intensity of the messages informing Alice of the diagnosis of a contact.
\end{itemize}
Whilst in protocol 2:
\begin{itemize}
    \item Grace learns the same as in protocol 1.
    \item Henry learns the same as in protocol 1 plus who has downloaded the app.
\end{itemize}

\noindent We now consider the cases of collusion. In the first protocol:
\begin{itemize}
    \item If Grace colludes with Mary they learn what Alice is being informed of.
    \item If Grace colludes with Henry they can learn how many people Bob has been in contact with and, psuedonymously, for how long and when.
    \item If Mary colludes with Henry they may be able to figure out who Bob has been in contact with (by matching messages they exchange with Grace on the basis of timing).
    \item If Grace colludes with snoopers then snoopers are enabled to track people.
    \item If Henry colludes with snoopers then he can learn who Bob has been in contact with.
\end{itemize}
However even if Grace, Henry and Mary all collude in the first protocol or if Grace and Henry collude in the second, they would still learn no more than what the government learns in BlueTrace. In particular, no information is leaked about contact between two people neither of which has been diagnosed.

\subsubsection{Correctness}

Due to information about a diagnoses having to go through Henry it is not possible for Bob to lie and say he has a diagnosis in order to cause panic. After Bob's diagnosis Alice will be informed of having had a contact with a diagnosed person, virtually instantaneously in protocol one and within thirty minutes in protocol two.

In these message based protocols there is an advantage in that the message can be used to convey auxiliary information, such as an estimate of the likelihood Alice has been infected. This would be computed by Grace and could be based on how many times she has been in contact with Bob, for how long, at what estimated distance and when the contact happened in relation to when Bob is estimated to have been contagious.

\section{A Set-Based Protocol: Enhanced Privacy from the Government using Additively Homomorphic Encryption} \label{sec:she protocol}

This section introduces our second approach to private contact tracing. At a high-level, our idea is to supplement the UIDs exchanged between users who engage in close-contact encounters with an additional 4~bit secret value, a one time pad. Every day, each user will interact with the government by uploading a list of tuples composed from every UID sent during an encounter in the last $N$ days and a corresponding 4~bit infection status indicator. If a user is infected then they will upload a 4~bit value equal to the secret value from their initial encounter, otherwise will upload a random other value.

Each user holds a public key pair for an additively homomorphic encryption scheme that they generate at the start of the protocol. Users periodically check their infection status by sending their public key alongside the list of newly encountered UIDs and the encryption of the corresponding secret 4~bit values, to the government. The government encrypts the corresponding 4~bit values uploaded by each of the UID owners using the public key and then computes the equality over the encrypted pairs. All of the encrypted equalities, each equivalent to learning an infection status, are returned to the user.

The additively homomorphic contact tracing protocol we propose can be split into the following three stages:
\begin{enumerate}
    \item Close-contact interactions are recorded.
    \item Each user periodically sends their record of interactions and their infection status to the government. 
    \item The government is periodically queried to learn of potentially infectious interactions.
\end{enumerate}

In more detail, stage 1 occurs whenever two people come into close contact with each other as defined in Section \ref{sec:Assumptions}. Stage 2, which occurs periodically such as every 24 hours, occurs when users upload pseudonym records of their close contact interactions to the government. Finally in stage 3, which also occurs periodically, each user learns whether any of their recorded interactions could have been infectious.

\begin{figure}
    \centering
    \includegraphics[scale=0.7]{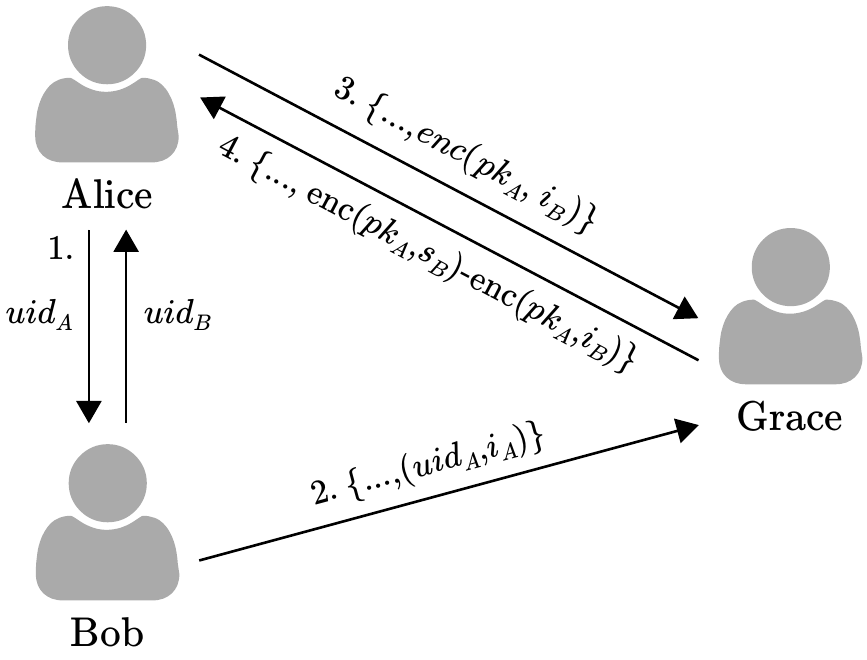}
    \caption{Overview of our set based contact tracing protocol.}
    \label{fig:set_based_protocol_1}
\end{figure}

\paragraph{Stage 1: Contact Recording}
The first stage of our protocol involves an arbitrary pair of users Alice and Bob that each hold a UID $uid \in \{0,1\}^{128}$ and a secret share $s \in \{0,1\}^4$ for the current period. Alice and Bob run the following protocol:

\begin{enumerate}

    \item Let $(uid_A, s_{A})$ and $(uid_B, s_{B})$ denote the random UID and share tuples that Alice and Bob, respectively, hold for the time period in which they meet.
    
    \item Alice sends Bob $(uid_A, s_{A})$ and Bob sends Alice $(uid_B, s_{B})$.
    
    \item Alice and Bob each add their individual UIDs, share values, and a timestamp $t$ and to the received tuple which is then added to the set of tuples they hold for all their interactions in the last $N$ days. In this instance:
    \begin{itemize}
    
    \item Alice holds the set $contact_A = \{\dots, (uid_B, s_{B}, uid_A, s_{A}, t_{B})\}$.
    
    \item Bob holds the set $contact_B = \{\dots, (uid_A, s_{A}, uid_B, s_{B}, t_{A})\}$.
    
    \end{itemize}
\end{enumerate}    

\paragraph{Stage 2: Status Reporting}

The second stage of our protocol involves a user of the system, Alice, and the government or service provider G. Alice will upload the set of tuples which comprises all of the UIDs she has communicated to another user during that day and a corresponding 4~bit infection status value.  
 
Using the $contact_A$ set described in Stage 1, then if Alice is known to be infectious she will upload the set of tuples for the last day comprising all of her personal UID values and ``secret'' shares from $contact_A$. If Alice is not infectious then she sends the same UID values but instead each with a random, other 4~bit value. G then takes the union of all such sets sent by all users. In more detail: 

\begin{enumerate}

    \item Each user $K$ constructs a status update set and sends it to G. The status update set comprises tuples of all of the UID values communicated in Stage 1 during the last day and the corresponding 4~bit infection status. If the user is infected then the infection status value $i_k$ is set to the ``secret'' share value $s_{1,K}$ from each interaction, otherwise a random other 4~bit value $i_K \xleftarrow{\$}\{0,\ldots,15\} \setminus s_{1,K}$ is chosen.
    
    The sets that Alice and Bob construct are given below. Here $t$ denotes the day the update set corresponds to i.e. the current day.
    
    $$\mathit{status\text{-}update}_{A,t} = \{\dots, (uid_A, i_A)\}$$
    $$\mathit{status\text{-}update}_{B,t} = \{\dots, (uid_B, i_B)\}$$
    
    \item At the end of this stage, G holds the status sets for all users for the last $N$ days.

    
    
    
    
\end{enumerate}

\paragraph{Stage 3: Infection Learning}
The third stage of our protocol involves a user of the system, Alice, and the government or service provider G. Alice will learn her infection risk by sending her public key, her set of encountered UIDs and the additively homomorphic encryption of the corresponding secret 4~bit values to the government. The government encrypts the latest 4~bit values uploaded by each of the encountered UIDs using Alice's public key and then computes the equality over the encrypted pairs. All of the encrypted inequalities, each equivalent to learning of a contact risking infection, are returned to the Alice. Alice finally decrypts the inequalities to learn if she has had an encounter that could have risked infection.

Alice holds her public key pair $(pk_A, sk_A)$ and her set of contact recordings $contact_A$. The government holds the set of all interactions $\mathit{interactions}_G$. The Stage 3 protocol is as follows: 

\begin{enumerate}
    \item Alice builds a set $\mathit{interacted\text{-}UIDs_A}$ comprising of all the additively homomorphic encryptions of the secrets shared in all Stage 1 encounters in the last $N$ days. Alice sends her public key $pk_A$ and the set  $\mathit{interacted\text{-}UIDs_A}$ to the government. Assuming Bob was her last interaction, the set is below, we denote the additively homomorphic encryption scheme by $\mathit{enc}$\footnote{It is important that order is preserved between this set and the status update sets sent by each user.}:
$$\mathit{interacted\text{-}UIDs_A} = \{\dots,  \mathit{enc}(pk_A,i_B)\}$$ 
    \item Alice sends $\mathit{interacted\text{-}UIDs_A}$ to the government. 
    
    \item For UID a user has encountered for which the government has received a infected status update from during Stage 2 of the protocol, it encrypts the corresponding status using Alice's public key $pk_A$. Next using the additively homomorphic properties of the underlying encryption scheme, the government is able to compute the (encrypted) inequality between all of the encrypted user status indicators  and the secret share values uploaded by Alice. If the government does not hold a given UID, or the status has not been updated for a ``long'' time, it is ignored. Finally, the government returns all of the encrypted status indicators to Alice. 
    
    \item Alice decrypts all of the ciphertexts sent to her by the government and learns whether she may be infected or not. Any ciphertext decryption that has the value 0 indicates that one of Alice's contacts has reported a positive infection status. 
    
\end{enumerate}

To compute the inequalities in Step 3 we can use any additive homomorphic encryption scheme, such as Paillier \cite{DBLP:conf/eurocrypt/Paillier99}, as described in Section \ref{sec:additive encryption}.

\subsection{Privacy of Our Set-Based Contact Tracing Method} \label{sec:privacy of set based method}
Here we briefly consider the privacy of our set-based contact tracing protocol in relation to the terms introduced in Section \ref{sec:privacy_terms}.

\paragraph{Privacy from Snoopers}
Compared to the Singaporean TraceTogether solution which we use as a baseline, our set-based protocol has equivalent privacy from snoopers. In particular, privacy from snoopers is essentially determined by our ``pseudonym change strategy'' of time-limited pseudonyms (UIDs) which remains unchanged from the TraceTogether design. 

\paragraph{Privacy from Contacts}
Each user only learns how many of their interactions in the last $N$ days exposed them to an individual who has reported an infection. Compared to TraceTogether this reveals more as there users only learns exactly what the government wants them to about infectious contacts. Our protocol, however, benefits from not depending on government policy and treats all users equally, infected or not.

\paragraph{Privacy from the Government}

Our set-based protocol offers user-controlled infection status disclosure. Uniquely, to the best of our knowledge, our protocol is the first to offer privacy from the government that is not dependent on infection status. Regardless of their infection status, all users report the pseudonymous identities they have exchanged and request infection indication values from the pseudonyms they have encountered. This does reveal more to the government than the TraceTogether solution which only discloses user behaviour for infected individuals, however our protocol never reveals to the government whether any user is infected or not --- which we believe to be a far more important privacy property.

Whilst this extra leakage may seem counter-intuitive, there are some unique advantages to our approach. Our set-based protocol ensures that only the user who may be at risk learns that one of their interactions could have caused them to become infected. The responsibility is then with the individual to get tested, self-isolate or seek treatment depending on their circumstances. In addition, the government can build a graph of pseuodonymous interaction that could allow it to model the impact of social distancing policies and in-turn make more accurate predictions about the spread of the virus and the implied demand for care resources. 

\section{Overhead/Performance} \label{sec:overhead}

In this section we consider the overhead and performance of the protocols we propose in this work. We look at the local storage requirements on the user and the overhead that is incurred. To approximate the overhead and performance of our protocols we make the following assumptions.

We note, in this version of the paper, we only intend to give an approximation of the overhead and performance of our methods.

Let $N=14$ days, and assume there are 10,000,000 users and that each had 100 close contact encounters each day\footnote{14 days is the current medically advised time period to consider. Considering 10,000,000 users is equivalent to using the system across an entire major city (e.g. London). Our assumption of 100 close encounters a day is our own estimate.}, we also assume that 50,000 people are newly infected each day\footnote{This is more than ten times the current rate of infection in the United Kingdom as of April 7th}. 

\subsection{Messaged-based protocols}

\subsubsection{User requirements}

\paragraph{Setup}
Having installed the app the user must generate and store two seeds $s$ and $t$. Both will take up 128 bits so between them that is 32 Bytes. The user must also encrypt their seeds and send the result to Mary or Henry (depending on which message protocol is used), this can be contained in a single cipher-text which will be no more than half a kilobyte.

\paragraph{Interactions}
The user must generate a new UID for every time period, this is one call to $\mathit{CTR}$. The parties broadcast a 128bit UID, this uses 16 of the available 20 Bytes in a BLE beacon so fits easily. The user must also store every UID they receive for a period of 14 days. Assuming 100 interactions a day this is 1400 UIDs together with a timestamp so that it is known when to stop storing them. A UID and timestamp together will require no more than 26 Bytes, so the total required storage will be about 36 kB.

\paragraph{Upon Diagnosis}
Upon being diagnosed the user uploads all the UIDs they have recently seen that is an upload of 36kB.

\paragraph{Receiving Updates}
The user must collect messages sent by Grace including the dummy messages. Assuming a message every 30 minutes on average this is 48 messages a day. If the messages consist of a single indicator bit then they are the size of the counter $n_A$ i.e. a few bytes, more informative messages would have to be bigger. Even with a fairly long messages, a few bytes say (much more information that this would probably de-anonymise the diagnosed person), the communication cost is going to be dominated by the meta data of a connection to Mary. Thus we expect the main communication cost to be the metadata associated with receiving 48 messages a day.

\subsubsection{Server requirements}

\paragraph{Henry as health care provider}
Henry must receive and forward a 36kB message for each diagnosed individual. He doesn't need to store anything or engage in any other steps in this role.

\paragraph{Mary or Henry as message relay}
Mary in protocol 1 or Henry in protocol 2 must receive setup details from every new user and forward them together with a pseudonym to Grace. This will mostly consist of a single cipher-text and so the communication depends on the scheme but wouldn't be much more than half a kilobyte in and out. For every user this party will also have to store the pseudonyms, which for 10,000,000 users could be a mere 3 bytes each, together with the identity of each person. This could be only 20 Bytes per user i.e. 200 MB total. For each of these users all messages from Grace will have to be forwarded by this party, assuming a message every 30 minutes this is 480,000,000 messages a day. Each message with consist of a pseudonym, a counter and a masked message, which could be only 10 Bytes. Thus even if the messages can't be delivered immediately (because the user is unavailable) and must be stored for a day on average this will require about 5GB of storage. This 5GB is also the amount of communication in and out each day, this should be fairly evenly spread. on top of this there will also be however much metadata is required for sending 480,000,000 messages each day, we expect this would be of the order of tens of gigabytes.

\paragraph{Grace}
Grace must send the above 480,000,000 ten byte messages everyday, this is 5GB but as they are all going to the same recipient their should be no extra meta data required. Grace must also receive 36kB of data for every newly diagnosed individual, with 50,000 individuals infected each day this is about 1.8GB a day. To store every message that is waiting to be sent in the next 30 minutes Grace would only need about 100MB of storage. Grace must store two keys, a pseudonym and a counter for each user amounting to around 32MB. In order to perform a lookup on a UID received from Henry, Grace will also want to maintain a lookup table, this will have to hold every UID used in the last 14 days, that is 14 times 48 times 10,000,000 or about 7 billion entries each returning a pseudonym of 3 Bytes. This will require tens possibly hundreds of gigabytes of storage depending on implementation. A call shall be made to this lookup table every time a UID is queried i.e. 70,000,000 times a day. This will require a lot of computation.

\subsection{Set-Based protocol}

We consider the different phases of the protocol in turn.

\paragraph{Stage 1: Contact reporting}

In each interaction a user sends $uid \in \{0,1\}^{128}$ and $s \in \{0,1\}^4$. Assuming 100 interactions a day this means each user must transfer no more than 1.65kB of data per day. Moreover, locally they must store their contact set. This logs all interactions for the last 14 days. Each element comprises of 2 UIDs, 2 shares and a timestamp. Assuming a timestamp requires 10 Bytes each element of the contact set requires 43 Bytes meaning the whole set requires approximately 60 kB of local storage by the user.

\paragraph{Stage 2: Status Reporting}

In this stage the user constructs and sends their status update set for that day to G. There are 100 elements in the status set (100 interactions per day), each comprising of a UID, $uid \in \{0,1\}^{128}$ and an indicator $i \in \{0,1\}^4$. Consequently the user must transfer approximately 1.7kB of data to the government every day. 

The government must store these sets for all users for the previous $N$ days. Assuming 10,000,000 users (the system is somehow split into regions) over 14 days this will require 230 GB. 

\paragraph{Stage 3: Infection Learning}

In this stage the user uses an additively homomorphic encryption scheme to learn if they have had an infected encounter. We base our results on using the Paillier encryption scheme implementation from \cite{DBLP:journals/iacr/JostLMS15}. We assume key size of 2048 bits, which is sufficient and recommended according to NIST \cite{NIST}. Using the results from \cite{DBLP:journals/iacr/JostLMS15} this means the user can do 89,483 encryptions per second requiring 5.66 seconds of pre-compute time. Therefore it will take less than a second for the user to encrypt all the 4 bit values in this stage along with the 5 seconds of pre-compute time. Given the user must only do this once a day, this is very reasonable. Moreover, the user must only transfer over the encrypted values of the indicators to the government for the last $N$ days, and not the UIDs they correspond to. Under our assumptions the set comprises of 1,400 2048-bit encryptions meaning a message size of 0.72 MB.\\  

As can be seen from the above analysis, the burden on the user is relatively small which is important if this method is to be widely adopted.

\begin{figure}\label{table:eID_comparison}
    \centering
    \begin{tabular}{ |p{2.7cm}||p{1.3cm}|p{1.3cm}|p{1.3cm}| }
     \hline
      & Stage 1 & Stage 2 & Stage 3  \\
     \hline
     User Storage & 60kB & -      & -                 \\
     User Transfer & -       & 1.7kB   & 0.36 MB           \\
     Govt Storage & -       & 230GB   & -        \\
     User Enc Time & -        &  -     &  7s               \\

     \hline
    \end{tabular}
    \caption{A summary of the approximate overheads per day for our protocol presented in Section \ref{sec:she protocol}, based on the assumptions given at the start of this section.}
    \label{fig:overhead}
\end{figure}
\section{Conclusion}

In this work we have proposed two methods that allow for privacy preserving contact tracing. Our first method expands on current implemented methods (e.g. TraceTogether) for private messaging and our second leverages the ability of the parties to exchange ``secrets" at the time of interaction and additively homomorphic encryption to solve the problem. Both our methods provide privacy for a user who is diagnosed with the virus, something that is not accounted for in existing methods. Moreover, in Section \ref{sec:overhead} we show how our methods are inexpensive and efficient, thus we believe they are of practical use in for the current pandemic and future situations where contact tracing is needed.

\bibliographystyle{plain}
\bibliography{biblio}

\end{document}